\documentstyle[12pt,fleqn]{article}
\def\singlespace {\smallskipamount=3.75pt plus1pt minus1pt
                  \medskipamount=7.5pt plus2pt minus2pt
                  \bigskipamount=15pt plus4pt minus4pt
                  \normalbaselineskip=15pt plus0pt minus0pt
                  \normallineskip=1pt
                  \normallineskiplimit=0pt
                  \jot=3.75pt
                  {\def\smallskip {\vskip\smallskipamount}}
                  {\def\medskip   {\vskip\medskipamount}}
                  {\def\bigskip   {\vskip\bigskipamount}}
                  {\setbox\strutbox=\hbox{\vrule 
                    height10.5pt depth4.5pt width 0pt}}
                  \parskip 7.5pt
                  \normalbaselines}
\def\middlespace {\smallskipamount=5.625pt plus1.5pt minus1.5pt
                  \medskipamount=11.25pt plus3pt minus3pt
                  \bigskipamount=22.5pt plus6pt minus6pt
                  \normalbaselineskip=22.5pt plus0pt minus0pt
                  \normallineskip=1pt
                  \normallineskiplimit=0pt
                  \jot=5.625pt
                  {\def\smallskip {\vskip\smallskipamount}}
                  {\def\medskip   {\vskip\medskipamount}}
                  {\def\bigskip   {\vskip\bigskipamount}}
                  {\setbox\strutbox=\hbox{\vrule 
                    height15.75pt depth6.75pt width 0pt}}
                  \parskip 11.25pt
                  \normalbaselines}
\def\doublespace {\smallskipamount=7.5pt plus2pt minus2pt
                  \medskipamount=15pt plus4pt minus4pt
                  \bigskipamount=30pt plus8pt minus8pt
                  \normalbaselineskip=30pt plus0pt minus0pt
                  \normallineskip=2pt
                  \normallineskiplimit=0pt
                  \jot=7.5pt
                  {\def\smallskip {\vskip\smallskipamount}}
                  {\def\medskip   {\vskip\medskipamount}}
                  {\def\bigskip   {\vskip\bigskipamount}}
                  {\setbox\strutbox=\hbox{\vrule 
                    height21.0pt depth9.0pt width 0pt}}
                  \parskip 15.0pt
                  \normalbaselines}

\include{dspace12}
\def\be{\begin{equation}}
\def\ee{\end{equation}}
\def\bea{\begin{eqnarray}}
\def\eea{\end{eqnarray}}
\def\nn{\nonumber}
\def\th{\theta}

\def\lt{\left}
\def\rt{\right}
\def\sect #1{\setcounter{equation}{0}}

\begin{document}
\middlespace
\vspace{0.5in}
\begin{center}
{\LARGE {Janis-Newman-Winicour and Wyman solutions are the same\footnotemark[1]}}
\end{center}
\footnotetext[1]{This paper is dedicated to the memory of Professor Nathan Rosen.}
\vspace{0.5in}
\vspace{12pt}
\begin{center}
{\large{ 
K. S. Virbhadra\\
Theoretical Astrophysics Group\\
Tata Institute of Fundamental Research\\
Homi Bhabha Road, Colaba, Mumbai 400005, India.\\
}}
\end{center}
\vspace{0.8in}

\begin{abstract}
We show that  the well-known  most general static and  spherically symmetric
exact solution to the Einstein-massless scalar equations given by 
Wyman is the same as one found by Janis, Newman and Winicour 
several years ago. We obtain the energy associated with  this spacetime 
and find that the total energy for the case of the purely scalar field 
is zero.
\end{abstract}
\vspace{0.2in}
\begin{center}
{\it To appear in  Int. J. Mod. Phys. A  }
\end{center}
\newpage
Even  before the general theory of relativity was proposed,  scalar
field has been conjectured to give rise to the long-range gravitational
fields\cite{conj}.
Several theories involving scalar fields have been suggested
\cite{theories}.
The subject of scalar fields  minimally as well as conformally coupled
to gravitation has fascinated many researchers' minds (\cite{Scalar}
-\cite{GHS91}).
Since the  last few years there has been a growing interest in studying
the gravitational collapse of scalar fields and the nature of singularities
in the Einstein-massless scalar (EMS) theory
(see Ref. \cite{sing} and references therein).
There has been considerable interest in obtaining solutions to
the EMS as  well as the Einstein-massless
conformal scalar equations.
Janis, Newman and Winicour (JNW) \cite{JNW68} obtained static
and spherically symmetric exact solution to the EMS equations. 
Wyman\cite{Wym81}  further obtained a static spherically symmetric
exact solution  to the EMS equations. There is no reference to
the JNW solution in his paper. Agnese and La Camera\cite{AL85} expressed
the Wyman solution in a more convenient form. 
Roberts\cite{Rob93} showed that the most general static spherically symmetic
solution to the EMS equations (with zero cosmological constant)
is asymptotically flat and this is the Wyman solution.
The Wyman solution is well-known in the literature\cite{WymCit}.In the present
note we show that the Wyman solution is the same as the JNW
solution, which was obtained almost twelve years ago.
We further obtain the total energy associated with this spacetime.
We use  geometrized units   and follow the
convention that Latin (Greek)   indices take values $0\ldots3$ ($1\ldots3$).

The EMS field equations are 
\be
R_{ij}\ -\ \frac{1}{2}\ R \ g_{ij}\ =\ 8 \pi \ S_{ij}\ ,
\ee
where $S_{ij}$, the energy-momentum tensor of the massless scalar field, is
given by
\be
S_{ij}\ =\ \Phi_{,i}\ \Phi_{,j}\ -\ \frac{1}{2}\ g_{ij}\ g^{ab}\ \Phi_{,a}\
          \Phi_{,b}\ ,
\ee
and 
\be
\Phi_{,i}^{\ ;i}\ =\ 0 .
\ee
 $\Phi$ stands for the massless scalar field. $R_{ij}$ is
the Ricci tensor and $R$ is the Ricci scalar.
Equation $(1)$ with Eq. $(2)$ can be expressed as
\be
R_{ij}\ =\ 8 \pi \ \Phi_{,i}\ \Phi_{,j} .
\ee
JNW\cite{JNW68} obtained  static and spherically symmetric  exact solution 
to the EMS equations, which is given by  the line element
\be
ds^2\ =\ {\lt( \frac{1-\frac{a_-}{R}}{1+\frac{a_+}{R}}
\rt)}^{1/{\mu}} dt^2 - {\lt( \frac{1-\frac{a_-}{R}}{1+\frac{a_+}{R}}
\rt)}^{-1/{\mu}} dR^2 
-  \lt(1-\frac{a_-}{R}\rt)^{1-1/{\mu}}
        \lt(1+\frac{a_+}{R}\rt)^{1+1/{\mu}}  R^2 d\Omega,
\ee
with
\be
d\Omega = d\th^2 + \sin^2\th d\phi^2,
\ee
and the scalar field 
\be
\Phi\ =\ \frac{\sigma}{\mu}
\ln \lt(
\frac{1-\frac{a_-}{R}}{1+\frac{a_+}{R}}
\rt) ,
\ee
where
\be
a_{\pm} = r_0 (\mu \pm 1)/2
\ee
and
\be
\mu = \sqrt{1+16 \pi {\sigma}^2} .
\ee
$r_0$ and $\sigma$ are two constant parameters in the solution.
$r_0=0$ gives the flat spacetime, whereas $\sigma =0$ gives the
line element
\be
ds^2 = \lt(1+\frac{r_0}{R}\rt)^{-1} dt^2 - \lt(1+\frac{r_0}{R}\rt) dR^2
     - \lt(1+\frac{r_0}{R}\rt)^2 R^ 2 d\Omega ,
\ee
which obviously represents the Schwarschild metric. The JNW
solution can be put in a more convenient form,  in coordinates 
$t,r,\th,\phi$,  by the line element
\be
ds^2 = \lt(1-\frac{b}{r}\rt)^{\gamma} dt^2 
      - \lt(1-\frac{b}{r}\rt)^{-\gamma} dr^2
      - \lt(1-\frac{b}{r}\rt)^{1-\gamma} r^2 d\Omega
\ee
and the scalar field
\be
\Phi = \frac{q}{b\sqrt{4\pi}} \ln\lt(1-\frac{b}{r}\rt),
\ee
where
\bea
\gamma &=& \frac{2m}{b}, \nn\\
b &=& 2 \sqrt{m^2+q^2}.
\eea
$m$ and $q$ are constant parameters. The radial coordinates $R$
and $r$ are related through
\be
r = R + a_{+}
\ee
and the spacetime parameters $r_0,\sigma$ and $m,q$ are related through
\bea
r_0 &=& 2 m , \nn\\
\sigma &=& \frac{1}{\sqrt{16\pi}} \frac{q}{m}.
\eea
The solution to the EMS equations, expressed by Eqs. $(11)-(13)$,
is exactly the well-known Wyman solution (see   Ref. \cite{AL85}). Thus,
Wyman did not obtain a  new solution, but he rediscovered the
JNW solution independently. The JNW solution, in $t,r,\th,\phi$
coordinates, has a curvature singularity at  $r=b$.  Garfinkle,
Horowitz and Strominger \cite{GHS91} obtained a nice form of charged dilaton
black hole solution, characterized by mass, charge, and a coupling
parameter (which controls the strength of the coupling of the dilaton
to the Maxwell field). A particular solution of this (putting
the electric charge parameter zero) yields the JNW (Wyman) solution.
This fact  is not noticed in their paper.

It is of interest to obtain the energy associated with the JNW spacetime.
The energy-momentum localization  has been a ``recalcitrant''
problem since the outset of the general theory of relativity.
Though, several energy-momentum complexes have been shown to  give the same 
and reasonable result (local values) for any Kerr-Schild class metric as well as 
for the Einstein-Rosen spacetime, the subject of energy-momentum
localization is still debatable (see Ref. \cite{Ksv} and references therein). However, 
the total energy
of an asymptotically flat spacetime is unanimously accepted.
Using the Einstein energy-momentum complex we first obtain 
an energy expression  for a general nonstatic spherically symmetric
metric and then we calculate the total energy associated with the JNW 
spacetime. A general nonstatic spherically symmetric line
element is
\be
ds^2\ = B dt^2 - A dr^2 - D r^2 \lt(d\theta^2 + \sin^2\theta d\phi^2\rt),
\ee
where $B = B\lt(r,t\rt), A = A\lt(r,t\rt), D = D\lt(r,t\rt)$.
The Einstein energy-momentum complex (see Refs. \cite{Ksv}) is
\be
\Theta_i^{\ k} =  \frac{1}{16 \pi} H^{\ kl}_{i \ \ ,l} ,
\ee
where
\be
H_i^{\ kl}\  =\ - H_i^{\ lk}\ =\  \frac{g_{in}}{\sqrt{-g}}
         \lt[-g \lt( g^{kn} g^{lm} - g^{ln} g^{km}\rt)\rt]_{,m} \ .
\ee
The Einstein energy-momentum complex  satisfies the local conservation laws
\be
\frac{\partial \Theta_i^{\ k}}{\partial x^k} = 0 ,
\ee
where
\be
\Theta_i^{\ k} = \sqrt{-g} \lt(T_i^{\ k} + \vartheta_i^{\ k}\rt).
\ee
$T_i^{\ k}$ is the symmetric energy-momentum tensor due to matter and all
nongravitational fields and $\vartheta_i^{\ k}$ is the energy-momentum 
pseudotensor due to the gravitational field only.
The energy and momentum components are given by
\be
P_i\ =\ \frac{1}{16 \pi} \ \int\int\ H_i^{0 \alpha} \ n_{\alpha}\ dS  ,
\ee
where  $dS$ is the infinitesimal surface element and  $n_{\alpha}$ is the
outward unit normal vector. $P_0$ stands for the energy and $P_{\alpha}$
stand for the linear momentum components.
 As it is known that the energy-momentum
complexes give meaningful result only if calculations are
carried out in quasi-Minkowskian coordinates, we transform the
line element in coordinates $t,x,y,z$ (according to $ x = r
\sin\theta \cos\phi, y = r \sin\theta \sin\phi, z = r
\cos\theta$). Then we calculate the required components of
$H_i^{\ kl}$ and these are
\bea
H_0^{\ 01}\ = \ 2 \sqrt{\frac{B}{A}} \  \frac{x\lt(A-D-D'r\rt)}{r^2},\nn\\
H_0^{\ 02}\ = \ 2 \sqrt{\frac{B}{A}} \  \frac{y\lt(A-D-D'r\rt)}{r^2},\nn\\
H_0^{\ 03}\ = \ 2 \sqrt{\frac{B}{A}} \  \frac{z\lt(A-D-D'r\rt)}{r^2},
\eea
where the prime denotes the partial derivative with respect to
the radial coordinate $r$. Using the above in $(21)$ we obtain
the  energy
\be
E = \ \frac{1}{2}\  \sqrt{\frac{B}{A}} \  r \lt(A-D-D'r\rt) .
\ee
Substituting $B = A^{-1} = {\lt(1-b/r\rt)}^{\gamma}$ and $D =  
{\lt(1-b/r\rt)}^{1-\gamma}$ we obtain  the  energy  associated 
with the JNW spacetime.
\be
E_{total}\ =\ m.
\ee
The total energy of the JNW spacetime is given by the parameter $m$. We have 
repeated these calculations using other energy-momentum
complexes and have found the same result as given by Eq. ($24$).
The total energy of  a  purely scalar field (i.e. for 
$m=0$) is zero. It is of interest to investigate whether
or not this is true in general, i.e. for any purely massless
scalar field.

\begin{flushleft}
{\large Acknowledgements}\\
Thanks are due to Professor P. C. Vaidya for a careful reading of the manuscript. 
\end{flushleft}
\newpage


\begin{thebibliography}{99}
\setlength{\parskip}{0.32ex}

\bibitem{conj}
    M. Abraham, {\em Jahrb. Radioakt. Electronik} {\bf 11}, 470 (1914);
    O. Bergmann,   {\em Am. J. Phys.} {\bf 24}, 38 (1956);
    S. Weinberg,  {\em Gravitation and cosmology: principles
          and applications of the general  theory of relativity
          (John Wiely \& Sons, NY, 1972)}, p.157.
\bibitem{theories}
     C. H. Brans  and  R. H.  Dicke,  {\em  Phys. Rev.} {\bf 124}, 925 (1961);
     C. G. Callan, S.  Coleman  and R. Jackiw,   {\em Ann. Phys.
            (N.Y.)} {\bf 59}, 42 (1970);
     J. H. Horne and  G. T. Horowitz,  {\em Phys. Rev.} {\bf D46}, 1340  (1992);
     H. Yilmaz,    {\em Il Nuovo Cimento} {\bf 107B}, 941 (1992);
     I. Peterson,   {\em Science News} {\bf 146}, 376 (1994).   
\bibitem{Scalar}
    R. Penny,   {\em Phys. Rev.}  {\bf 174},   1578 (1968);
    J. D. Bekenstein,   {\em  Ann.  Phys.}  {\bf 91},  75 (1975);
    S. S. Bayin,  F. I.  Cooperstock   and V.   Faraoni,
             {\em Astrophy. J.} {\bf 428}, 439 (1994);
    K. S. Virbhadra and J. C. Parikh,   {\em Phys. Lett.} {\bf B331}, 302 (1994);
                 Erratum : {\em Phys. Lett.} {\bf B340}, 265 (1994);
    K. S. Virbhadra,  gr-qc/9408035, {\em  Pramana-J.Phys.}  {\bf 44}, 317 (1995);
    A. E. Mayo and  J. D. Bekenstein,   {\em Phys. Rev.} {\bf D54}, 5059 (1996);
    C. Martinez  and J.  Zanelli,   Phys. Rev.  {\bf D54}, 3830 (1996);
    J. D. Bekenstein, {\em Black hole hair: twenty-five years}, gr-qc/9605059 (1996).
\bibitem{sing}
    E. Malec,  {\em Selfgravitating nonlinear scalar fields}, Commun. Math. 
               Phys., (1996)  (submitted).
\bibitem{JNW68}
      A. I. Janis, E. T.  Newman and J.  Winicour,   {\em  Phys. 
           Rev. Lett.}   {\bf 20},  878 (1968).
\bibitem{Wym81}
      M. Wyman,   {\em Phys. Rev.} {\bf D24}, 839 (1981).
\bibitem{AL85}
    A. G. Agnese   and M. La Camera,  {\em Phys. Rev.} {\bf D31}, 1280 (1985).
\bibitem{Rob93}
   M. D. Roberts,  Astrophys. \& Space Sc. {\bf 200}, 331 (1993).
\bibitem{WymCit}
 T. Papacostas,  {\em J. Math. Phys.} {\bf 32}, 2468 (1991);
D. V. Galtsov and  B. C. Xanathopoulos,  {\em J. Math. Phys.} {\bf 33}, 273 (1992);
K. Schmoltzi and T. Schucker,  {\em Phys. Lett. A} {\bf 161}, 212 (1991);
B. C. Xanthopoulos and T. E.  Dialynas,   {\em J. Math. Phys.} {\bf 33}, 1463 (1992);
P. Jetzer and D. Scialom,  {\em Phys. Lett. A} {\bf 169}, 12 (1992);
J. Z. Li,   {\em J. Math. Phys.} {\bf 33}, 3506 (1992);
A. Hardell  and  H. Dehnen,  {\em Gen. Relativ. Gravit.} {\bf 25}, 1165 (1993);
V. Husain, E. A.  Martinez  and D. Nunez,   {\em Phys.Rev.} {\bf D50}, 3783 (1994);
H. P. Deoliveira,  {\em J. Math. Phys.} {\bf 36}, 2988 (1995);
Y. J. Kiem  and D.  Park,   {\em Phys. Rev.} {\bf D53}, 747 (1996);
J. M. Salim and  S. L. Sautu,   {\em Class. Quantum Grav.} {\bf 13}, 353 (1996).
\bibitem{GHS91}
    D. Garfinkle, G. T.  Horowitz   and A.   Strominger,   {\em Phys.  Rev.}
       {\bf D43}, 3140 (1991); Erratum :  {\em Phys. Rev.} {\bf D45}, 3888 (1992).
\bibitem{Ksv}
     K. S. Virbhadra,  Phys. Rev. {\bf D41}, 1086 (1990);
                       Phys. Rev. {\bf D42}, 1066 (1990);
                       Phys. Rev. {\bf D42}, 2919 (1990);
    N. Rosen and K. S.  Virbhadra,   Gen. Relativ. Grav. {\bf 25}, 429 (1993);
    K. S. Virbhadra and J. C. Parikh,   Phys. Lett. {\bf B317}, 312 (1993);
    K. S. Virbhadra,  Pramana- J. Phys. {\bf 45}, 215 (1995);
    J. M. Aguirregabiria, A. Chamorro  and  K. S. Virbhadra, 
                            Gen. Relativ. \& Gravit. {\bf 28}, 1393 (1996).

\end{thebibliography}
\end{document}